\documentclass[a4paper,10pt,onecolumn,oneside,notitlepage,final]{article}
\usepackage{graphics}

\begin{document}
\title{\bf Two-loop Radiative Corrections to Neutrino Masses \\ 
        in $SU(3)_L \times U(1)_N$ Gauge Models}
\author{Teruyuki Kitabayashi
            \footnote{A talk given at {\it Post Summer Institute 2000 
            on Neutrino Physics}, August 21-24, 2000, Fuji-Yoshida, Japan.}
            \footnote{E-mail:teruyuki@post.kek.jp}
        \vspace{2mm}
        \\
        {\small \sl Accelerator Engineering Center,}\\ 
        {\small \sl Mitsubishi Electric System \& Service Engineering Co.Ltd.,} \\
        {\small \sl 2-8-8 Umezono, Tsukuba, Ibaraki 305-0045, Japan.}\\
        }
\maketitle

\begin{abstract}
We have constructed two $SU(3)_L \times U(1)_N$ gauge models with the $L^\prime = L_e-L_\mu-L_\tau$ symmetry, which accommodate tiny neutrino masses generated by one-loop and two-loop radiative effects. The heavy neutral leptons and heavy charged leptons are employed to specify the lepton triplets in $SU(3)_L$, respectively, accompanied by ($\phi^0$, $\phi^-$, $h^+$) and ($\phi^+$, $\phi^0$, $h^+$), where $\phi$ stands for the standard Higgs scalar and $h^+$ is a key ingredient for radiative mechanisms. From our numerical calculations, we find that both our models are relevant to yield the VO solution to the solar neutrino problem. 
\end{abstract}

\section{Introduction}
\hspace{4mm}
The Super-Kamiokande experiments on atmospheric neutrino oscillations provide evidence for tiny neutrino masses and their mixings \cite{Kamiokande,RecentSK}. Solar neutrinos are observed to be also oscillating \cite{Solar}. To generate such tiny neutrino masses, two main theoretical mechanisms have been proposed: one is the seesaw mechanism \cite{Seesaw} and the other is the radiative mechanism \cite{1-loop,2-loop,Radiative,ZeeType}. In the radiative mechanism proposed by Zee \cite{1-loop}, a new singly charged $SU(2)$-singlet Higgs scalar, $h^+$, was introduced into the $SU(2)_L \times U(1)_Y$ standard model and neutrino masses were generated as one-loop radiative corrections via the $h^+$ coupling to $\ell_L\nu_L$. After this work, Zee and Babu studied two-loop radiative mechanism \cite{2-loop}. In two-loop mechanism, one more doubly charged $SU(2)$-singlet Higgs scalar $k^{++}$ was added to the standard model and tiny neutrino masses arose from two-loop radiative effects generated by $\ell_R\ell_R k^{++}$.

In this talk, I report my recent works \cite{Neutral,Charged} about generation of tiny neutrino masses and oscillations in $SU(3)_L \times U(1)_N$ models \cite{SU3U1}. The observed structure in atmospheric and solar neutrino oscillations is considered to be based on the presence of the $L_e-L_\mu-L_\tau$ conservation \cite{Lprime,LprimeTwoLoop}, which ensures maximal solar neutrino mixing and the hierarchy of $\Delta m^2_{atm}\gg\Delta m^2_\odot$ although $\Delta m^2_\odot$=0 and on the generic smallness of one-loop effects over two-loop effects to yield tiny amount of $\Delta m^2_\odot$ realizing $\Delta m^2_{atm}\gg\Delta m^2_\odot$ ($\neq$ 0) \cite{2vs1Loop}.  To induce maximal atmospheric neutrino mixing needs other physical origin, which is ascribed to approximate degeneracy between masses of newly introduced heavy leptons.  Altogether, we finally obtain (approximate) bimaximal neutrino mixing scheme for neutrino oscillations \cite{Mixing,NearlyBiMaximal}. 

\section{Models}
\hspace{4mm}
The radiative neutrino mechanism has been discussed to consistently yield one-loop radiative neutrino masses in various $SU(3)_L \times U(1)_N$ models \cite{Okam99}. To further discuss two-loop effects in these models, we have constructed two $SU(3)_L \times U(1)_N$ models: {\it the heavy neutral leptons model} \cite{Neutral} and {\it the heavy charged lepton model} \cite{Charged}.

\subsection{Heavy Neutral Lepton Model}
\hspace{4mm}
In this model, quantum numbers of the leptons and the Higgs scalars are summarized as follows: 
\\
\\
\textbf{Leptons}
\begin{equation}
\psi^{i=1,2,3}_L = 
    \left(\nu^i, \ell^i, \omega^{i} \right)_L^T   
                     : \left( \textbf{3},-\frac{1}{3}    \right), \quad
    \ell^  {1,2,3}_L : \left( \textbf{1},-1         \right), \quad  
    \omega^{1,2,3}_R : \left( \textbf{1}, 0   \right),
\label{Eq:N_leptons}
\end{equation}
\textbf{Higgs Scalars}
\begin{eqnarray}
\rho &=&  \left( \rho^+, \rho^0, \overline\rho^+ \right)^T
               : \left( \textbf{3}, \frac{2}{3} \right), \quad
\rho^\prime = \left(\rho^{\prime +}, \rho^{\prime 0}, \overline\rho^{\prime +} \right)^T
               : \left( \textbf{3}, \frac{2}{3} \right),
\nonumber \\
\eta &=&  \left(\eta^0, \eta^-, \overline\eta^0 \right)^T
               : \left( \textbf{3}, -\frac{1}{3} \right), \quad
\chi =    \left(\overline\chi^0, \chi^-, \chi^0 \right)^T 
               : \left( \textbf{3}, -\frac{1}{3} \right),
\nonumber \\
&&  k^+ : (\textbf{1},1),
\label{Eq:N_HiggsScalars}
\end{eqnarray}
where $\omega^{0i}$ stand for three heavy neutral leptons, $k^+$ is an extra scalar that plays a role of Babu's $k^{++}$ and  values in the parentheses specify quantum numbers based on the $(SU(3)_L,U(1)_N)$-symmetry. The Zee's scalar, $h^+$, corresponds to $\overline\rho^{\prime +}$.  Three Higgs scalars have the following vacuum expectation values (VEV): $<0|\eta|0>=(v_\eta,0,0)^T, <0|\rho|0>=(0,v_\rho,0)^T, <0|\chi|0>=(0,0,v_\chi)^T$, where this orthogonal choice of VEV's is to be guaranteed by the $\eta\rho\chi$-term in Eq.(\ref{Eq:N_V0_Vb}), and the remainder, $\rho^\prime$, acquire no VEV for model to be consistently described \cite{VEVofRhoPrime}. We also impose the 
$L^\prime = L_e-L_\mu-L_\tau$ conservation \cite{Lprime} on our interactions to reproduce the observed atmospheric neutrino oscillations. For the charge assignment, see Table \ref{TabLnumberNM}.
\begin{table}[h]
\caption{
    \label{TabLnumberNM}
    $L$ and $L^\prime$ quantum numbers in heavy neutral lepton model.}
\begin{center}
\begin{tabular}{ccccc}
    \hline
        Fields & $\eta,\rho,\rho^\prime,\chi$ & $\psi_L^1,\ell_R^1,\omega_R^1$
               & $\psi_L^{2,3},\ell_R^{2,3},\omega_R^{2,3}$ & $k^+$ \\
    \hline
        $L$        & 0 & 1 &  1 & 2 \\    
    \hline
        $L^\prime$ & 0 & 1 & -1 & 2 \\
    \hline
\end{tabular}
\end{center}
\end{table}

The Higgs interactions are given by self-Hermitian terms composed of at most four scalar fields and by two types of non self-Hermitian Higgs potentials, $L^\prime$-conserving potential ($V_0$) and $L^\prime$-violating potential ($V_b$): 
\begin{eqnarray}
V_0 &=& \lambda_0 \epsilon^{\alpha\beta\gamma} \eta_\alpha \rho_\beta \chi_\gamma
      + \lambda_1 (\rho^\dagger \eta       )(\chi^\dagger \rho^\prime)
      + \lambda_2 (\rho^\dagger \rho^\prime)(\chi^\dagger \eta       )
      + (h.c.),
\nonumber \\
V_b &=& \mu_b \rho^{\prime\dagger} \eta k^+ + (h.c.),
\label{Eq:N_V0_Vb}
\end{eqnarray}
where $\lambda_{0,1,2}$ are the coupling constants and $\mu_b$ denotes the $L^\prime$-breaking mass scale. If there is only the $V_0$, the eigenvalues of the neutrino mass matrix are given by $0$ and $\pm m_\nu$ ($m_\nu$: neutrino mass). From these eigenvalues, we can describe only atmospheric neutrino oscillations. However, if the $V_b$ also exists, we can realize a two-loop radiative mechanism and we successfully obtain both atmospheric \cite{Kamiokande,RecentSK} and solar \cite{Solar} neutrino oscillations. This is why we have introduced the $V_b$.

The Yukawa interactions relevant for the neutrino mass generation are given by the following lagrangian:
\begin{eqnarray}
-\mathcal{L}_Y &=&
     \frac{1}{2}\epsilon^{\alpha\beta\gamma}\sum_{i=2,3}f_{[1i]}
     \overline{\left(\psi_{\alpha L}^1 \right)^c} \psi_{\beta L}^i \rho^\prime_\gamma
   + \sum_{i=1,2,3} \overline{\psi_L^i}
     \left( f_{\ell}^i \rho \ell_R^i + f_\omega^i \chi \omega_R^i \right)
\nonumber \\
 &&+ \sum_{i,j=2,3} f_k^{ij} \overline{(\ell_R^i)^c} \omega_R^j k^+ + (h.c.),
\label{Eq:N_Yukawa}
\end{eqnarray}
where $f$'s are Yukawa couplings with the relation $f_{[ij]}=-f_{[ji]}$ demanded by the Fermi statistics. 

\subsection{Heavy Charged Lepton Model}
\hspace{4mm}
In this model, the leptons and the Higgs scalars are summarized as follows:
\\
\\
\textbf{Leptons}
\begin{equation}
\psi^{i=1,2,3}_L = \left( \nu^i, \ell^i, \kappa^{i} \right)_L^T   
                     : \left( \textbf{3}, 0 \right), \quad
    \ell^  {1,2,3}_R : \left( \textbf{1},-1 \right), \quad  
    \kappa^{1,2,3}_R : \left( \textbf{1}, 1 \right),
\label{Eq:C_leptons}
\end{equation}
\textbf{Higgs scalars}
\begin{eqnarray}
&&  \eta = \left(\eta^0, \eta^-, \eta^+    \right)^T : \left( \textbf{3},0  \right), \quad
    \rho = \left(\rho^+, \rho^0, \rho^{++} \right)^T : \left( \textbf{3},1  \right),
\nonumber \\
&&  \chi = \left(\chi^-, \chi^{--}, \chi^0 \right)^T : \left( \textbf{3},-1 \right),
\quad
    k^{++} : (\textbf{1},2),
\label{Eq:C_HiggsScalars}
\end{eqnarray}
where $\kappa^{+i}$ stand for three heavy charged leptons and $k^{++}$ is an extra scalar that plays a role of Babu's $k^{++}$. The Zee's scalar, $h^+$, corresponds to $\eta^+$.  The Higgs scalars have the following vacuum expectation values (VEV): $<0|\eta|0>=(v_\eta,0,0)^T, <0|\rho|0>=(0,v_\rho,0)^T, <0|\chi|0>=(0,0,v_\chi)^T$. In the same way as in the heavy neutral lepton model, we also impose the $L^\prime$ conservation in this model. Listed 
in Table \ref{TabLnumberCM} are various $L$- and $L'$-quauntum  numbers.
\begin{table}[h]
\caption{
    \label{TabLnumberCM}
    $L$ and $L^\prime$ quantum number in heavy charged lepton model.}
\begin{center}
\begin{tabular}{ccccc}
    \hline
        Fields & $\eta,\rho,\chi$ & $\psi_L^1,\ell_R^1,\kappa_R^1$
               & $\psi_L^{2,3},\ell_R^{2,3},\kappa_R^{2,3}$ & $k^{++}$ \\
    \hline
        $L$        & 0 & 1 &  1 & 2 \\    
    \hline
        $L^\prime$ & 0 & 1 & -1 & 2 \\
    \hline
\end{tabular}
\end{center}
\end{table}

The Higgs interactions are also given by self-Hermitian terms composed of at most four scalar fields and by two types of non self-Hermitian Higgs potentials, $L^\prime$-conserving potential ($V_0$) and $L^\prime$-violating potential ($V_b$): 

\begin{eqnarray}
V_0 &=& \lambda_0 \epsilon^{\alpha\beta\gamma} \eta_\alpha \rho_\beta \chi_\gamma
      + \lambda_1 (\chi^\dagger \eta)(\rho^\dagger \eta)
      + (h.c.),
\nonumber \\
V_b &=& \mu_b \rho^\dagger \chi k^{++} + (h.c.),
\label{Eq:C_V0_Vb}
\end{eqnarray}
where $\lambda_{0,1}$ are the coupling constants and $\mu_b$ denotes the $L^\prime$-breaking mass scale again. 

The Yukawa interactions relevant for the neutrino mass generation are given by the following lagrangian:
\begin{eqnarray}
-{\mathcal{L}}_Y &=& 
     \frac{1}{2}\epsilon^{\alpha\beta\gamma}\sum_{i=2,3}f_{[1i]}
     \overline{\left(\psi_{\alpha L}^1 \right)^c} \psi_{\beta L}^i \eta_\gamma
   + \sum_{i=1,2,3} \overline{\psi_L^i}
     \left( f_\ell^i \rho \ell_R^i + f_\kappa^i \chi \kappa_R^i \right)
\nonumber \\
&& + f_k^{11} \overline{(\kappa_R^1)^c} (k^{++})^\dagger \kappa_R^1
   + (h.c.).
\label{Eq:C_Yukawa}
\end{eqnarray}
Note that the possible interactions of $\Sigma_{i,j=2,3} f_k^{ij} \overline{(\ell_R^i)^c} k^{++} \ell_R^j$ are forbidden by the $L$-conservation.  

\section{Neutrino masses and oscillations}
\hspace{4mm}
Now, we can discuss how radiative corrections induce neutrino masses in our models.

\subsection{Heavy neutral lepton model}
\begin{figure}[t]
  \begin{center}
    \includegraphics*[30mm,200mm][190mm,275mm]{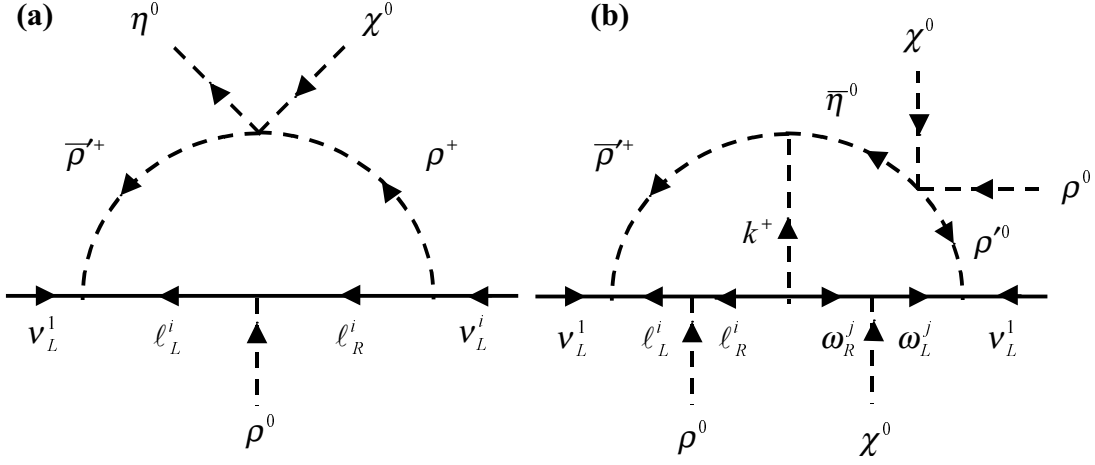}
    \caption{\label{fig1}
    Loop diagrams in the heavy neutral lepton model, 
    (a) an example of one-loop diagrams for $\nu^1-\nu^i~(i=2,3)$,
    (b) two-loop diagrams for $\nu^1-\nu^1$.}
  \end{center}
\end{figure}

\hspace{4mm}
In this model, we obtain that one-loop and two-loop diagrams, as shown in Fig.\ref{fig1}, correspond to the following interactions
\begin{eqnarray}
&&  \left( \eta^\dagger \psi_L^1     \right)
      \epsilon^{\alpha\beta\gamma} \rho_\alpha \chi_\beta \psi_{\gamma L}^{2,3}
+ \left( \eta^\dagger \psi_L^{2,3} \right)
      \epsilon^{\alpha\beta\gamma} \rho_\alpha \chi_\beta \psi_{\gamma L}^1,
\nonumber \\
&& \epsilon^{\alpha\beta\gamma} \epsilon^{\delta\epsilon\zeta}
    \psi_{\alpha L}^1 \rho_\beta \chi_\gamma \psi_{\delta L}^{2,3} \rho_\epsilon \chi_\zeta.
\label{Eq:N_Effective}
\end{eqnarray}
From one-loop diagrams, we can calculate the following Majorana masses to be:
\begin{eqnarray}
m_{1i}^{(1)} &=& f_{[1i]}
    \Biggl[ 
    \lambda_1 
    \frac{ m_{\ell i}^2 F 
    \left( m_{\ell i}^2,m_{\overline{\rho} \prime 0}^2,m_{\rho +}^2 \right)
         - m_{\ell 1}^2 F \left(m_{\ell 1}^2,m_{\overline{\rho} \prime 0}^2,m_{\rho +}^2
    \right)}
    {v_\rho^2}
\nonumber \\
&& +\lambda_2 
    \frac{ m_{\omega i}^2 F 
    \left( m_{\omega i}^2,m_{\rho \prime 0}^2,m_{\overline{\chi} 0}^2 \right)
       - m_{\omega 1}^2 F \left(m_{\omega 1}^2,m_{\rho \prime 0}^2,m_{\overline{\chi} 0}^2     \right)
    }
    {v_\chi^2}
    \Biggl] v_\eta v_\rho v_\chi,
\nonumber \\
F(x,y,z)&=&\frac{1}{16\pi^2}
    \Biggl[
     \frac{x\log x}{(x-y)(x-z)}+\frac{y\log y}{(y-x)(y-z)}
    +\frac{z\log z}{(z-y)(z-x)}\Biggl].
\label{Eq:N_m1_F}
\end{eqnarray}
In addition, from two-loop diagrams, the outline of its derivation is shown in the Appendix of \cite {Neutral}, we find
\begin{eqnarray}
m_{11}^{(2)}&=& -2 \sum_{i=2,3}\lambda_2 
              f_{[1i]}f_k^{ij}f_{[1j]} m_{\ell i} m_{\omega j} \mu_b v_\rho v_\chi I,
\nonumber \\
I &=&\frac{1}{m_\eta^2-m_{\rho \prime}^2 }
     \left[ J(m_\eta^2)-J(m_{\rho \prime}^2) \right],
\nonumber \\
J(m^2)&=&\frac{1}{m_k^2}G(m_\ell^2, m_{\rho \prime}^2, m_k^2)
                        G(m_\omega^2, m^2, m_k^2),
\nonumber \\
G(x,y,z)&=&\frac{1}{16\pi^2} \frac{x\log(x/z)-y\log(y/z)}{x-y},
\label{Eq:N_m2_I_J}
\end{eqnarray}
where $m_{\ell i}(\equiv f_{\ell i}^i v_\rho)$ and $m_{\omega i}(\equiv f_{\omega i}^i) v_\chi$ are the mass of the $i$-th charged lepton and the mass of the $i$-th heavy neutral lepton respectively, and masses of Higgs scalars are denoted by the subscripts in terms of their fields. 

Now, we obtain the following neutrino mass matrix:
\begin{equation}
M_\nu=
    \left(
    \begin{array}{ccc}
         m_{11}^{(2)} & m_{12}^{(1)} & m_{13}^{(1)} \\
         m_{12}^{(1)} & 0            & 0            \\
         m_{13}^{(1)} & 0            & 0            \\
    \end{array}
    \right),
\label{Eq:N_Mnu}
\end{equation}
from which we find
\begin{equation}
\Delta m_{atm}^2 = m_{12}^{(1)2}+m_{13}^{(1)2} =m_\nu^2, \quad
\Delta m_\odot^2 = 2m_\nu |m_{11}^{(2)}|.
\label{Eq:N_Delta_m}
\end{equation}
where we have used the relation of $|m_{11}^{(2)}| \ll |m_{12}^{(1)}|,|m_{13}^{(1)}|$.

In order to see whether our model gives the compatible description of neutrino oscillations with the observed data, we make the following assumptions on relevant free parameters:
\begin{enumerate}
\item 
Since $v_{\eta,\rho}$ is related to masses of weak bosons proportional to $\sqrt{v^2_\eta+v^2_\rho}$, we require $\sqrt{v^2_\eta+v^2_\rho}$=$v_{weak}$, from which ($v_\eta$, $v_\rho$) $\sim$ ($v_{weak}/10$, $v_{weak}$) are taken, where $v_{weak}$ = $( 2{\sqrt 2}G_F)^{-1/2}$ = 174 GeV, 
\item 
Since $v_\chi$ is a source of masses for heavy charged leptons and also of masses for exotic quarks and exotic gauge bosons, we use $v_\chi$ $\gg$ $v_{weak}$, from which $v_\chi \sim 10v_{weak}$ is taken,
\item 
The masses of the Higgs bosons, $\eta$ and $\rho^\prime$, are set to be $m_\eta \sim m_{\rho \prime} \sim v_{weak}$,
\item 
The masses of the Higgs bosons, $k^+$ and $\chi$, and of the heavy charged leptons, $\omega^i$ ($i$ = 1,2,3), are assigned to be larger values as $m_k$ $\sim$ $m_\chi$ $\sim$ $v_\chi$ and $m_{\omega 2, \omega 3}$ $\sim$ $ev_\chi$ supplemented by 10\% mass difference between $\omega^1$ and $\omega^{2,3}$, {\it i.e.} $m_{\omega 1}$ $\sim$ $0.9m_{\omega 2, \omega 3}$, where $e$ stands for the electromagnetic coupling,
\item 
The $L$-violating couplings of $f_{[1i]}$ ($i$ = 2,3) are determined by $\Delta m^2_{atm}$, where $f_{[1i]}$ $\sim$ $10^{-7}$ is to be taken,
\item 
The $L^\prime$-violating scale of $\mu_b$ is suppressed as $ev_\chi$,
\item 
The $L$- and $L^\prime$-conserving couplings accompany no suppression factor and are set to be of order 1 as $f_k^{ij} \sim \lambda_1 \sim  \lambda_2 \sim 1$. 
\end{enumerate}
These values are tabulated in Table \ref{TabParamNM}. 
\begin{table}[h]
 \caption{\label{TabParamNM}Model parameters in heavy neutral lepton model, where masses are given in the unit of $v_{weak}=(2\sqrt{2}G_F)^{-1/2}=174$ GeV and $e$ stands for the electromagnetic coupling.}
 \begin{center}
  \begin{tabular}{ccccccccccc}
    \hline
     $v_\eta$    & $v_\rho$   & $v_\chi$   & $m_{k,\chi}$ & $m_{\eta,\rho}$ & $m_{\omega 1}$ & $m_{\omega 2,3}$ & $\mu_b$&
     $\lambda_{1,2}$ & $f_k$ & $f_{ij}$\\
    \hline
       1/10      & 1     & 10   & 10       & 1       & $9e$       &$10e$&       $10e$    & 1     & 1    & $ 10^{-7}$ \\
    \hline
  \end{tabular}
 \end{center}
\end{table}

From numerical analysis, we obtain that $\Delta m_{atm}^2 =3 \times 10^{-3}$ eV$^2$, $\Delta m_\odot^2 = 2 \times 10^{-10}$ eV$^2$ and $\sin 2\vartheta$ = 0.97. We can see, these values lie in the allowed region of the observed solar neutrino oscillations relevant to the VO solution \cite{VOComment}. 

\subsection{Heavy charged lepton model}
\begin{figure}[t]
 \begin{center}
    \includegraphics*[30mm,200mm][190mm,275mm]{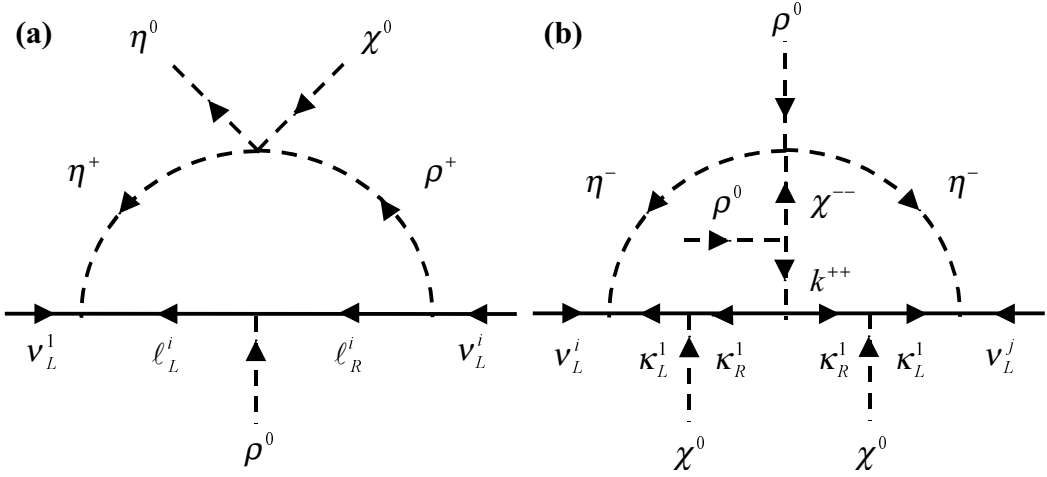}
 \end{center}
 \caption{\label{fig2}
Loop diagrams in the heavy charged lepton model, 
(a) an example of one-loop diagrams for $\nu^1-\nu^i~(i=2,3)$,
(b) two-loop diagrams for $\nu^i-\nu^j~(i,j=2,3)$.}
\end{figure}

\hspace{4mm}
From similar discussions in the previous subsection, we obtain one-loop and two-loop diagrams, as shown in Fig.\ref{fig2}, which correspond to 
\begin{eqnarray}
&& \left( \eta^\dagger \psi_L^1     \right)
      \epsilon^{\alpha\beta\gamma} \rho_\alpha \chi_\beta \psi_{\gamma L}^{2,3}
+ \left( \eta^\dagger \psi_L^{2,3} \right)
      \epsilon^{\alpha\beta\gamma} \rho_\alpha \chi_\beta \psi_{\gamma L}^1,
\nonumber \\
&&\epsilon^{\alpha\beta\gamma} \epsilon^{\delta\epsilon\zeta}
    \psi_{\alpha L}^i \rho_\beta \chi_\gamma \psi_{\delta L}^j \rho_\epsilon \chi_\zeta.
\label{Eq:C_Effective}
\end{eqnarray}
From one-loop diagrams, we obtain the following Majorana masses,
\begin{eqnarray}
m_{1i}^{(1)} &=& f_{[1i]}
    \lambda_1 
    \Biggl[ 
    \frac{
        m_{\ell i}^2 F \left(m_{\ell i}^2,m_{\eta +}^2,m_{\rho +}^2 \right)
      - m_{\ell 1}^2 F \left(m_{\ell 1}^2,m_{\eta +}^2,m_{\rho +}^2 \right)
    }
    {v_\rho^2}
\nonumber \\
&& +\frac{
        m_{\kappa i}^2 F \left(m_{\kappa i}^2,m_{\eta -}^2,m_{\chi -}^2 \right)
      - m_{\kappa 1}^2 F \left(m_{\kappa 1}^2,m_{\eta -}^2,m_{\chi -}^2 \right)
    }
    {v_\chi^2}
    \Biggl] v_\eta v_\rho v_\chi.
\label{Eq:C_m1}
\end{eqnarray}
And, from two-loop diagrams, we find
\begin{eqnarray}
m_{ij}^{(2)} &=& -2\lambda_1f_{[1i]}f_k^{11}f_{[1j]}  \mu_bm_{\kappa 1}^2 v^2_\rho I,
\nonumber \\
I &=& \frac{1}{m_\chi^2-m_k^2} \left[ J(m^2_\chi)-J(m^2_k) \right],
\nonumber \\
J(m^2) &=& \frac{1}{m^2}G^2(m_\kappa^2,m_\eta^2,m^2).
\label{Eq:C_I_J}
\end{eqnarray}
\\
I show the outline of its derivation in the Appendix of \cite {Charged}.

Now, we obtain the following neutrino mass matrix:
\begin{equation}
M_\nu =   
    \left(
    \begin{array}{ccc}
         0 & m_{12}^{(1)} & m_{13}^{(1)} \\
         m_{12}^{(1)} & m_{22}^{(2)} & m_{23}^{(2)} \\
         m_{13}^{(1)} & m_{23}^{(2)} & m_{33}^{(2)} \\
    \end{array}
    \right),
\label{Eq:C_Mnu}
\end{equation}
from which we find
\begin{eqnarray}
&& \Delta m_{atm}^2 = m_{12}^{(1)2}+m_{13}^{(1)2} \quad (=m_\nu^2), 
\nonumber \\
&& \Delta m_\odot^2 = 2 \left(  m_{22}^{(2)} \cos^2 \vartheta
                             + 2m_{23}^{(2)} \cos\vartheta \sin\vartheta
                             +  m_{33}^{(2)} \sin^2 \vartheta
                        \right) m_\nu,          
\label{Eq:C_Delta_m}
\end{eqnarray}
where $\cos\vartheta = m_{12}^{(1)}/m_\nu$ and $\sin\vartheta= m_{13}^{(1)}/m_\nu$.

In the same way as in the heavy neutral lepton model, we specify various parameters in this model with the following assumptions on relevant free parameters:
\begin{enumerate}
\item 
  ($v_\eta$, $v_\rho$) $\sim$ ($v_{weak}/10$, $v_{weak}$) with $\sqrt{v^2_\eta+v^2_\rho}$=$v_{weak}$, 
\item 
$v_\chi \sim 10v_{weak}$,
\item 
The masses of the Higgs bosons, $\eta$ and $\rho$, are set to be $m_\eta \sim m_\rho \sim v_{weak}$,
\item 
The masses of the Higgs bosons, $k^{++}$ and $\chi$, and of the heavy charged leptons, $\kappa^i$ ($i$ = 1,2,3), are assigned to be larger values, we take  $m_k$ $\sim$ $m_\chi$ $\sim$ $v_\chi$ with $m_{\kappa 1} \sim 0.9m_{\kappa 2, \kappa 3}$ where $m_{\kappa 2, \kappa 3}$ $\sim$ $ev_\chi$,
\item 
The $L$-violating couplings of $f_{[1i]}$ are also determined by $\Delta m^2_{atm}$, where $f_{[1i]}$ $\sim$ $10^{-7}$ is taken,
\item 
The $L^\prime$-violating scale of $\mu_b$ is suppressed as $ev_\chi$,
\item 
$f^k_{11}$ $\sim$ $\lambda_1$ $\sim$ 1.
\end{enumerate}
These values are tabulated in Table \ref{TabParamCM}.
 \begin{table}[h]
 \caption{\label{TabParamCM}Model parameters in heavy charged lepton model, where masses are given in the unit of $v_{weak}=(2\sqrt{2}G_F)^{-1/2}=174$ GeV and $e$ stands for the electromagnetic coupling.}
 \begin{center}
  \begin{tabular}{ccccccccccc}
    \hline
    $v_\eta$    & $v_\rho$   & $v_\chi$   & $m_{k,\chi}$ & $m_{\eta,\rho}$ & $m_{\kappa 1}$ & $m_{\kappa 2,3}$ & $\mu_b$&
     $\lambda_1,$ & $f_k$ & $f_{ij}$\\
    \hline
       1/10      & 1     & 10   & 10       & 1       & $9e$       &$10e$&
       $10e$    & 1     & 1    & $ 10^{-7}$ \\
    \hline
  \end{tabular}
 \end{center}
\end{table}

From numerical analysis, we obtain that $\Delta m_{atm}^2 = 3 \times 10^{-3}$ eV$^2$, $\Delta m_\odot^2 = 2.8 \times 10^{-10}$ eV$^2$ and $\sin 2\vartheta = 0.97$. These values lie in the allowed region of the observed solar neutrino oscillations relevant to the VO solution again \cite{VOComment}. 

\section{\label{SecSummary}Summary}
\hspace{4mm}
We have constructed two $SU(3)_L \times U(1)_N$ gauge models, which provide the radiatively generated neutrino masses and the observed neutrino oscillations. In the first model, Zee's scalar, $h^+$, is placed in ($\phi^+$, $\phi^0$, $h^+$), which requires the heavy neutral leptons $\omega^{0i}$ to specify the lepton triplets in $SU(3)_L$. On the other hand, in the second model, $h^+$ is accompanied by ($\phi^0$, $\phi^-$) to form ($\phi^0$, $\phi^-$, $h^+$), which requires the heavy charged leptons $\kappa^{+i}$. To account for the observed hierarchy between $\Delta m^2_{atm}$ $\gg$ $\Delta m^2_\odot$, we have used the $L^\prime=L_e-L_\mu-L_\tau$ symmetry and the generic smallness of one-loop radiative effects over two-loop radiative effects.

From numerical estimates, we obtain that $\Delta m^2_{atm}$ = $3 \times 10^{-3}$ eV$^2$ with $\sin 2\vartheta$ = 0.97 and $\Delta m_\odot^2$ = $2 \times 10^{-10}$ eV$^2$ in the heavy neutral lepton model, also $\Delta m^2_{atm}$ = $3 \times 10^{-3}$ eV$^2$ with $\sin 2\vartheta$ = 0.97 and $\Delta m_\odot^2$ = $2.8 \times 10^{-10}$ eV$^2$ in the heavy charged lepton model. Thus, both our models are relevant to yield the VO solution to the solar neutrino problem.

To find other possibilities of yielding tiny neutrino masses and oscillations in the $SU(3)_L \times U(1)_N$ models, especially the $(\nu,e,E^-)$ model \cite{HeavyE} that seems to yield the LOW solution, is our future study.

\begin{center}
{\bf Acknowledgments}
\end{center}

\hspace{4mm}
The author would like to thank Prof. M. Yasu\`{e} for valuable comments, helpful discussions and continuous encouragement. 



\begin{thebibliography}{99}
\bibitem{VOComment}
Although the recent report from Super-Kamiokande has presented the statement that the VO solution seems to be disfavored at the 95\% confidence level \cite{RecentSK}, it is stressed that this statement is not conclusive\cite{PSI2000}.  Therefore, the VO solution is still a possible solution to solar neutrino problem.
\end{thebibliography}
\end{document}